\documentclass[12pt,a4paper]{article}
\usepackage{amsfonts}
\usepackage{mathrsfs}  
\usepackage{epsfig}
\usepackage{axodraw}
\usepackage{graphicx}

\def\simlt{\stackrel{<}{{}_\sim}}

\begin{document}

\title{On the ground-state energy of a mixture of two
 different oppositely polarized fermionic gases}
\author{\em Piotr Chankowski and Jacek Wojtkiewicz\footnote{Emails:
    chank@fuw.edu.pl, wjacek@fuw.edu.pl} \\
Faculty of Physics, University of Warsaw,\\
Pasteura 5, 02-093 Warszawa, Poland
}
\maketitle
\abstract{We report the results of the computation of the order
  $(k_{\rm F}a_0)^2$ correction, where $k_{\rm F}=3\pi^2\rho$ is the 
  Fermi wave vector and $a_0$ the $s$-wave scattering length of  
  the repulsive interaction, to the ground-state energy of a
  mixture of oppositely polarized $N_a$ spin $1/2$ fermions $a$
  of masses $m_a$ and $N_b$ spin $1/2$ fermions $b$ of masses
  $m_b$ ($\rho=N/V$, $N=N_a+N_b$). It is shown that the results of
  the paper \cite{FraPil} in which the same correction has been
  computed entirely numerically, using a more traditional approach,
  can be easily and semianalytically reproduced using the effective
  field theory technique.
\vskip0.1cm

}

\newpage

\section{Introduction}
\label{sec:introd}

Whether the ferromagnetic behaviour of a gas of spin $1/2$ fermions (i.e.
the emergence of the so-called itinerant ferromagnetism) can be induced
by their repulsive spin-independent interaction is experimentally still
an open issue which is being studied by exploiting the upper branch of
the Feshbach resonance allowing to appropriately tune the interaction
strength of fermionic atoms of ultra-cold gases \cite{Jo,Lee}. Theoretical
studies of this problem have a long history. The classic mean-field
calculation \cite{Sto} predicts that the critical interaction strength
above which the ground state energy $E_\Omega$ of the polarized gas of
spin $1/2$ fermions is lower than that of the nonpolarized one is
$k_{\rm F}a_0=\pi/2$, where $k_{\rm F}=3\pi^2N/V$ is the gas Fermi vector
and $a_0$ is the $s$-wave scattering length characterizing the repulsive
interaction. More recently computations of the ground-state energy going
beyond the mean-field approximation \cite{Cond}, also ones exploiting the
quantum Monte Carlo simulations \cite{QMC10}, have resulted in a lower
critical value, $k_{\rm F}a_0\approx0.8$. This relatively large
critical interaction strength seems to be a source of considerable
difficulties in experimental observation of the effect \cite{Lee,Pek}.

Among different possible ways of favouring the appearance of
ferromagnetism the use of a mixture of oppositely polarized different
fermionic atomic gases (of different masses) has been proposed. The
computation of the ground-state energy of such a mixture in the
approximation going one order beyond the simple mean-field one, i.e. up
to terms of order $(k_{\rm F}a_0)^2$ has been undertaken in \cite{FraPil}
and a variety of possible phases of the system has been exhibited by a
detailed numerical study.

The order $(k_{\rm F}a_0)^2$ correction in the perturbative expansion of
the ground-state energy $E_\Omega$ of interacting fermionic system can be
computed either more traditionally, as in \cite{FraPil}, by the method
outlined in \cite{AbGoDz} which leads to rather complicated multiple
integrals which must be evaluated numerically, or using the effective
theory \cite{HamFur00}
(see also \cite{HAKOL,KolckiSka} for applications of this method to
many-body systems). The latter method is particularly well suited for
the case in which the interaction potential is not given explicitly but
is from the outset characterized only by the set of scattering lengths
$a_\ell$ and effective radii $r_\ell$, $\ell=0,1,\dots$ It allowed to
easily recover \cite{HamFur00} the classic order $(k_{\rm F}a_0)^2$ result
in the case of a unpolarized gas of identical fermions of arbitrary spin
and to extend it up to yet higher orders \cite{HamFur00,WeDrSch}. Very
recently we have used it \cite{PJ1} to obtain the order $(k_{\rm F}a_0)^2$
correction to the ground state energy of the diluted polarized gas of
identical spin 1/2 fermions, easily numerically recovering (and thereby
demonstrating its universality) the old analytic result of Kanno
\cite{KANNO} which has been obtained by the method of \cite{AbGoDz}
for the specific hard-core interaction potential.

It is a natural step to extend the effective theory approach to
the case of a diluted mixture of oppositely polarized fermions
of different masses. We present this extension in this paper.
It turns out that it reduces to only a small modification of the
computation done in \cite{PJ1} and as there, part of the
computations can be done analytically; the remaining integrals
are simple (compared to the ones done in \cite{FraPil}) and can
be easily evaluated numerically with the help of a three-line
Mathematica code using its standard built-in integration routines.
Moreover, the correctness of the computation is partially controlled
by the cancellation of ultraviolet divergences.
The computation is sketched in Section \ref{sec:comp}
and the comparison and the discussion are given in
Section \ref{sec:res}.

\section{Computation}
\label{sec:comp}

\noindent We consider a mixture of $N_a$ spin $1/2$ (nonrelativistic)
fermions of masses $m_a$ ($a$-fermions), all having spins up and $N_b$
spin $1/2$ fermions of masses $m_b$ having spins down ($b$-fermions),
enclosed in a box of volume $V$ and interacting with one another through
a spin-independent two-body short range repulsive potential. In
  the traditional language of quantum mechanics the Hamiltonian $H$
  of the system is of the form
  \begin{eqnarray}
    H=-{\hbar^2\over2m_a}\!\sum_{i_a=1}^{N_a}\!
    \mbox{\boldmath{$\nabla$}}^2_{i_a}-{\hbar^2\over2m_b}\!\sum_{i_b=1}^{N_b}\!
    \mbox{\boldmath{$\nabla$}}^2_{i_b}+\sum_{i_a,i_b}\!
    V_{\rm pot}(|\mathbf{r}_{i_a}-\mathbf{r}_{i_b}|)\nonumber\\
    +{1\over2}\!\sum_{i_a\neq j_a}\!
    V_{\rm pot}(|\mathbf{r}_{i_a}-\mathbf{r}_{j_a}|)+{1\over2}\!\sum_{i_b\neq j_b}\!
    V_{\rm pot}(|\mathbf{r}_{i_b}-\mathbf{r}_{j_b}|)~\!,\phantom{a}~\label{eqn:Hfund}
  \end{eqnarray}
  where $V_{\rm pot}(|\mathbf{r}|)$ is a repulsive, spin independent
  interaction potential and the wave function $\psi$ of the system,
  satisfying periodic boundary conditions in the box of volume
  $V=L\times L\times L$
  should be properly antisymmetrized in its $N_a$ arguments
  $(\mathbf{r}_{i_a},s_{i_a})$ and in its $N_b$ arguments
  $(\mathbf{r}_{i_b},s_{i_b})$, $s_{i_{a/b}}=\pm{1\over2}$. In the rest of
the paper the more convenient formalism of the second quantization is used.
Without loss of generality we assume that $N_a\geq N_b$ (the ratio
$m_b/m_a$ can be arbitrary).

If the gas of fermions is diluted, so
that the Fermi wave vector $k_{\rm F}=(3\pi^2N/V)^{1/3}$ (where $N=N_a+N_b$)
is sufficiently small, its ground-state energy $E_\Omega$ can be computed
using the effective theory approach \cite{HamFur00}. As follows from the
analysis done there, to obtain $E_\Omega$ up to the order $(k_{\rm F}a_0)^2$,
it is sufficient to restrict oneself to the lowest dimension interaction
operator, i.e. to consider the Hamiltonian of
form\footnote{Since in the considered system there are no $a$-fermions
  with spin down ($b$-fermions with spin up), the possible interactions of 
  $a$-fermions ($b$-fermions) between themselves do not play any role in
  determination of $E_\Omega$
  owing to the Pauli exclusion principle and the nonrelativistic character
  of the theory (impossibility of particle-antiparticle pair creation) and
  can, therefore, be omitted.}
\begin{eqnarray}
  H_{\rm eff}=H_0+V_{\rm int}=\sum_{\mathbf{p}}\left({\hbar^2\mathbf{p}^2\over2m_a}~\!
  a^\dagger_{\mathbf{p}} a_{\mathbf{p}}+{\hbar^2\mathbf{p}^2\over2m_b}~\!
  b^\dagger_{\mathbf{p}} b_{\mathbf{p}}\right)
  +{C_0\over V}\sum_{\mathbf{q}}\sum_{\mathbf{p}_1,\mathbf{p}_2}
  a^\dagger_{\mathbf{p}_1+\mathbf{q}}a_{\mathbf{p}_1}
  b^\dagger_{\mathbf{p}_2-\mathbf{q}}b_{\mathbf{p}_2}~\!,
  \label{eqn:effHamiltonian}
\end{eqnarray}
(the most general effective Hamiltonian has in principle infinitely many
operator structures of growing dimension \cite{HamFur00,HAKOL,KolckiSka}).
In contrast to the underlying ``fundamental'' Hamiltonian (\ref{eqn:Hfund}),
the effective one, (\ref{eqn:effHamiltonian}), is strictly local.
The zeroth and first order contributions to $E_\Omega$
\begin{eqnarray}
  E_\Omega^{(0)}+E_\Omega^{(1)}={V\over6\pi^2}~\!{3\over5}~\!{\hbar^2\over2}
  \left({p^5_{{\rm F}a}\over m_a}+{p^5_{{\rm F}b}\over m_b}\right)
  +VC_0~\!{p^3_{{\rm F}a}\over6\pi^2}~\!{p^3_{{\rm F}b}\over6\pi^2}~\!,
  \label{eqn:E0and1}
\end{eqnarray}
in which $p_{{\rm F}a/b}=(6\pi^2N_{a/b}/V)^{1/3}$ are the Fermi wave-vectors
of the $a$- and $b$-fermions, can be then immediately obtained
by applying to the Hamiltonian (\ref{eqn:effHamiltonian}) the ordinary
Rayleigh-Schr\"odinger expansion in conjunction with the standard
methods of second quantization \cite{FetWal,Feynman}. The coefficient
$C_0$ has to be related
to the $s$-wave scattering length $a_0$ which is extracted from the
expansion ($k=|\mathbf{k}|$)
\begin{eqnarray}
  f(k,\theta)=-a_0\left[1-ia_0k+\left({1\over2}~\!a_0r_0-a_0^2\right)k^2
    +\dots\right]-a_1^3k^2\cos\theta+\dots,\label{eqn:fScattExp}
\end{eqnarray}
of the amplitude of the elastic scattering of the $a$- and $b$-fermions with
the wave vectors $\mathbf{k}_a=\mathbf{k}$ and $\mathbf{k}_b=-\mathbf{k}$,
generated by the interaction $V_{\rm int}$ of (\ref{eqn:effHamiltonian}).
The amplitude $f(k,\theta)$ can in turn be obtained from the corresponding
$S$-matrix element $S_{\beta\alpha}$ computed in the second quantization
formalism with the help of the standard formula \cite{Weinb}
\begin{eqnarray}
  S_{\beta\alpha}=\langle\mathbf{k}^\prime_a,\mathbf{k}^\prime_b|
  {\rm T}\exp\!\left(-{i\over\hbar}\!\int_{-\infty}^\infty
  \!dt~\!V_{\rm int}^I(t)\right)\!|\mathbf{k}_a,\mathbf{k}_b\rangle
  \label{eqn:Smatrix}\\
  \equiv\delta_{\beta\alpha}-{i\over\hbar}~\!(2\pi)^4
  \delta^{(4)}(k_a^\prime+k_a^\prime-k_a-k_b)~\!{\cal A}~\!,~\!\nonumber
\end{eqnarray}
in which $V_{\rm int}^I(t)$
is the interaction picture counterpart of the interaction term of
(\ref{eqn:effHamiltonian}) written in the continuum normalization
\begin{eqnarray}
  V^I_{\rm int}(t)=C_0\int\!d^3\mathbf{x}~\!\psi^\dagger_a(t,\mathbf{x})~\!
  \psi_a(t,\mathbf{x})~\!
  \psi^\dagger_b(t,\mathbf{x})~\!\psi_b(t,\mathbf{x})~\!,
  \label{eqn:interaction}\\
  \psi_a(t,\mathbf{x})=\int\!{d^3\mathbf{k}\over(2\pi)^3}~\!
  e^{-i\omega^a_{\mathbf{k}}t+i\mathbf{k}\cdot\mathbf{x}}~\!a(\mathbf{k})~\!,
  \phantom{aaaaaaaaaaaa}\nonumber
\end{eqnarray}
etc. and  T is the symbol of the chronological ordering; employed
is also the ``four-vector'' notation in which
$k^0_{a/b}=\omega^{a/b}_{\mathbf{k}}\equiv\hbar\mathbf{k}^2/2m_{a/b}$.
The necessary rule is
\begin{eqnarray}
  f(k,\theta)=-{m_{\rm red}\over2\pi\hbar^2}~\!{\cal A}(k,\theta)~\!,
  \label{eqn:Rule}
\end{eqnarray}
where $m_{\rm red}\equiv m_am_b/(m_a+m_b)$ is the reduced mass of the
interacting fermions.
In the lowest order of the Dyson expansion of (\ref{eqn:Smatrix})
one readily finds (see e.g. \cite{HamFur00,HAKOL}) that
$C_0=(2\pi\hbar^2/m_{\rm red})a_0$. This allows to express (\ref{eqn:E0and1}) -
the first nontrivial approximation to the ground-state energy -
in terms of a physical quantity $a_0$. 

The local character of the interaction term of the Hamiltonian
(\ref{eqn:effHamiltonian}) results in ultraviolet divergences
in higher order corrections, both to the scattering amplitude ${\cal A}$
extracted from (\ref{eqn:Smatrix}) and to $E_\Omega$; the corrections
to the result (\ref{eqn:E0and1}) can be most conveniently computed
using the formula\footnote{The symbol T of the chronological ordering
  should not be confused with $T$ denoting time.}
\begin{eqnarray}
  \lim_{T\rightarrow\infty}\exp(-iT(E_\Omega-E_{\Omega_0})/\hbar)=
  \lim_{T\rightarrow\infty}\langle\Omega_0
  |{\rm T}\exp\!\left(-{i\over\hbar}\!\int_{-T/2}^{T/2}\!dt~\!
  V^I_{\rm int}(t)\right)\!|\Omega_0\rangle~\!.\label{eqn:basicFormula}
\end{eqnarray}
according to which $(E_\Omega-E_{\Omega_0})/V$ is directly given by $i\hbar$
times the sum of the momentum space connected
vacuum Feynman diagrams (the factor $(2\pi)^4\delta^{(4)}(0)$ arising in
evaluating connected vacuum diagrams in the position space is interpreted
as $VT$). The divergences, if regularized in the same way in evaluating
the formulae (\ref{eqn:Smatrix}) and (\ref{eqn:basicFormula}), disappear
from the result for $E_\Omega$ when $C_0$ and coefficients of other operator
structures of the effective Hamiltonian are in it consistently, order by
order, traded for the scattering lengths $a_\ell$ and the effective ranges
$r_\ell$ extracted from the computed scattering amplitude.

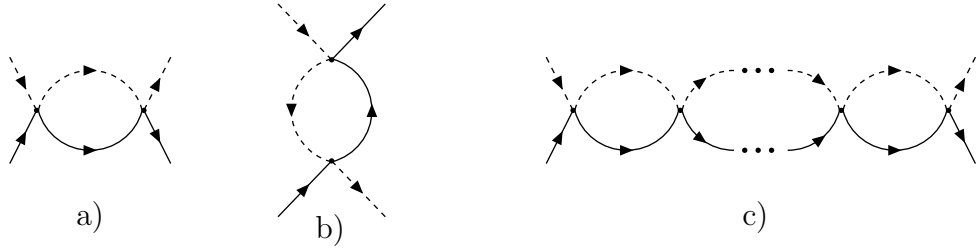
\begin{figure}[]
\begin{center}
\begin{picture}(400,80)(5,0)
\ArrowArc(30,45)(20,195,345)
\DashArrowArcn(30,35)(20,165,15){2}
\Vertex(10,40){1}
\Vertex(50,40){1}
\ArrowLine(0,20)(10,40)
\DashArrowLine(0,60)(10,40){2}
\ArrowLine(50,40)(60,20)
\DashArrowLine(50,40)(60,60){2}
\Text(30,0)[]{a)}
\ArrowArc(115,40)(20,285,75)
\DashArrowArc(125,40)(20,105,255){2}
\Vertex(120,21){1}
\Vertex(120,59){1}
\ArrowLine(100,0)(120,21)
\DashArrowLine(120,21)(140,0){2}
\DashArrowLine(100,80)(120,59){2}
\ArrowLine(120,59)(140,80)
\Text(120,-5)[]{b)}
\ArrowArc(230,45)(20,195,345)
\DashArrowArcn(230,35)(20,165,15){2}
\ArrowArc(270,45)(20,195,270)
\DashArrowArcn(270,35)(20,165,90){2}
\Vertex(274,55){1}
\Vertex(279,55){1}
\Vertex(284,55){1}
\Vertex(274,25){1}
\Vertex(279,25){1}
\Vertex(284,25){1}
\Text(279,0)[]{c)}
\ArrowArc(290,45)(20,270,345)
\DashArrowArcn(290,35)(20,90,15){2}
\ArrowArc(330,45)(20,195,345)
\DashArrowArcn(330,35)(20,165,15){2}
\Vertex(210,40){1}
\Vertex(250,40){1}
\Vertex(310,40){1}
\Vertex(350,40){1}
\ArrowLine(200,20)(210,40)
\DashArrowLine(200,60)(210,40){2}
\ArrowLine(350,40)(360,20)
\DashArrowLine(350,40)(360,60){2}
\end{picture}
\end{center}
\caption{Two one-loop Feynman
  diagrams representing the elastic scattering
  amplitude of the $a$-fermion (solid lines) on the $b$-fermion
  (dashed lines); the diagram b) vanishes.
  The ``sausage''-type diagrams c) originating in higher orders
  from the interaction term proportional to $C_0$.
  Time flows from the left to the right.}
\label{fig:ScatteringOneLoop}
\end{figure}

Here we regularize the divergences by cutting off all integrals over
the wave vectors $\mathbf{k}$ at the scale $\Lambda$. The limit
$\Lambda\rightarrow\infty$ will be taken after expressing $E_\Omega$
computed in terms of $a_\ell$ and $r_\ell$'s (as in \cite{PJ1} the
cancellation of the terms diverging as $\Lambda\rightarrow\infty$ will
serve as a partial check of the correctness of the calculation). Thus,
to obtain the complete
correction $E^{(2)}_\Omega$ to the result (\ref{eqn:E0and1}), $C_0$ in
$E^{(1)}_\Omega$ has to be expressed through $a_0$ up to one-loop order.
The interaction (\ref{eqn:interaction}) leads to two one-loop diagrams
shown in Figs. \ref{fig:ScatteringOneLoop}a and \ref{fig:ScatteringOneLoop}b,
representing scattering of $a$-fermions on $b$-fermions. The second one
vanishes, however, owing to the absence of antiparticles; moreover
it is easy to see that this interaction generates a whole class of diagrams
shown in Fig.~\ref{fig:ScatteringOneLoop}c which can be easily taken into
account.  Evaluating them  using the standard
Feynman rules \cite{FetWal} with the propagators
\begin{eqnarray}
  \langle{\rm void}|{\rm T}\psi_{a/b}(t,\mathbf{x})
  \psi^\dagger_{a/b}(t^\prime,\mathbf{x}^\prime)
  |{\rm void}\rangle=\int\!{d^3\mathbf{k}\over(2\pi)^3}~\!
  e^{i\mathbf{k}\cdot(\mathbf{x}-\mathbf{x}^\prime)}\!
  \int\!{d\omega\over2\pi}~\!{i~\!e^{-i\omega(t-t^\prime)}\over
    \omega-\omega^{a/b}_{\mathbf{k}}+i0}~\!,\nonumber
\end{eqnarray}
one obtains for the  scattering amplitude the expression\footnote{Upon
  integrating over frequencies with the help of the residue method the
  denominators of the propagators neatly combine so that the result depends
  only on $m_{\rm red}$.}
\begin{eqnarray}
  f(k,\theta)=
  -{m_{\rm red}\over2\pi\hbar^2}~\!C_0\left\{1
  +\left({C_0\over i\hbar}\right)\left({2m_{\rm red}\over i\hbar}~\!I_0\right)
  +\left({C_0\over i\hbar}\right)^2
  \left({2m_{\rm red}\over i\hbar}~\!I_0\right)^2+\dots\right\},\label{eqn:fAmp}
\end{eqnarray}
where 
\begin{eqnarray}
  I_0=\int\!{d^3\mathbf{q}\over(2\pi)^3}~\!
  {1\over\mathbf{q}^2-\mathbf{k}^2-i0}~\!.\label{eqn:I0integralDef}
\end{eqnarray}
The amplitude (\ref{eqn:fAmp}), supplemented in general with ($k$-dependent)
terms which come from diagrams generated by the other interactions
(omitted in (\ref{eqn:effHamiltonian})) of the
effective Hamiltonian should be matched onto the expansion
(\ref{eqn:fScattExp}).
The integral $I_0$ is divergent and requires regularization. Imposing the
UV cut-off $\Lambda$ on $q=|\mathbf{q}|$ one obtains ($k=|\mathbf{k}|$)
\begin{eqnarray}
I_0(k,\Lambda)={1\over4\pi^2}\!\int_0^\Lambda\!dq~\!q\left[{1\over q-k-i0}
+{1\over q+k+i0}\right]
={i\over4\pi}~\!k+{1\over2\pi^2}~\!\Lambda-{1\over2\pi^2}~\!
{k^2\over\Lambda}+\dots,\label{eqn:I0integralResult}
\end{eqnarray}
upon using the standard Sochocki formula $1/(x\pm i0)=P(1/x)\mp i\pi\delta(x)$
($P$ stands for principal value).
Inserting this into the formula (\ref{eqn:fAmp}) matched onto the
expansion (\ref{eqn:fScattExp}) and solving for  $C_0$ one finds
\begin{eqnarray}
C_0={2\pi\hbar^2\over m_{\rm red}}~\!a_0\left(1+{2\over\pi}~\!a_0
\Lambda+\dots\right).\label{eqn:C0determined}
\end{eqnarray}
\vskip0.2cm

The right hand side of the formula (\ref{eqn:basicFormula}) can be
evaluated using the standard rules of the many-body quantum field theory
(see e.g. \cite{FetWal}). Because $|\Omega_0\rangle$ is the lowest
energy state of $N_a$ free $a$-fermions and $N_b$ free $b$-fermions, in
the momentum space lines of Feynman diagrams correspond to the propagators 
\begin{eqnarray}
  i\tilde G^{(0)}_{a/b}(\omega,\mathbf{k})
  =i\left[{\theta(|\mathbf{k}|-p_{{\rm F}a/b})\over
  \omega-\omega^{a/b}_{\mathbf{k}}+i0}+{\theta(p_{{\rm F}a/b}-|\mathbf{k}|)\over
  \omega-\omega^{a/b}_{\mathbf{k}}-i0}\right].\label{eqn:MomSpaceProp}
\end{eqnarray}
and, to account for the normal ordered form of the interaction term in
(\ref{eqn:effHamiltonian}), one has only to add the rule \cite{FetWal}
that if a line originates from and ends up in one and the same vertex,
the propagator (\ref{eqn:MomSpaceProp}) corresponding to this line has
to be multiplied by $e^{i\omega\eta}$ with the limit $\eta\rightarrow0^+$
taken at the end.

\begin{figure}[]
\begin{center}
\begin{picture}(80,40)(5,0)
\CArc(20,35)(20,195,345)
\CArc(20,25)(20,15,165)
\DashCArc(60,35)(20,195,345){2}
\DashCArc(60,25)(20,15,165){2}
\Vertex(40,30){2}
%
\end{picture}
\end{center}
\caption{The effective theory connected vacuum diagram of order $C_0$
  reproducing the first order correction $E_\Omega^{(1)}$.
  Solid and dashed lines represent propagators of
  $a$- and $b$- fermions, respectively.}
\label{fig:VacuumDiagrSpinOneHalf}
\end{figure}
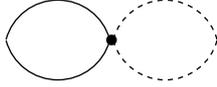

In the first order in $C_0$ there is only one connected vacuum
graph shown in Figure \ref{fig:VacuumDiagrSpinOneHalf} which
(evaluated in the position space) immediately gives
$(iG^{(0)}_{a/b}(0)$ are the  propagators (\ref{eqn:MomSpaceProp})
written in the position space)
\begin{eqnarray}
  T E^{(1)}_\Omega
  =C_0~\!VT~\!iG^{(0)}_a(0)~\!iG^{(0)}_a(0)=C_0~\!VT~\!{p^3_{{\rm F}a}\over6\pi^2}
  ~\!{p^3_{{\rm F}b}\over6\pi^2}~\!,\nonumber
\end{eqnarray}
which reproduces the result (\ref{eqn:E0and1}).
\vskip0.2cm

\begin{figure}[]
\begin{center}
\begin{picture}(80,40)(5,0)
\ArrowArc(30,20)(25,70,290)
\DashArrowArc(30,20)(25,290,70){2}
\DashArrowArc(50,20)(25,110,250){2}
\ArrowArc(50,20)(25,250,110)
\Vertex(40,-2.5){1}
\Vertex(40,42.5){1}
\Text(84,20)[]{$^k$}
\Text(64,20)[]{$^p$}
\Text(-10,20)[]{$^{k+q}$}
\Text(39,20)[]{$^{p-q}$}
\end{picture}
\end{center}
\caption{The only nonvanishing three-loop connected vacuum diagram
  contributing the order $(k_{\rm f}a_0)^2$ correction to the ground state
  energy of the diluted gas of the
  mixture of (spin $1/2$) $a$- and $b$-fermions. The
  two kinds of propagators differ by the values of the Fermi momenta;
  for definiteness it is assumed that $p_{{\rm F}a}\geq p_{{\rm F}b}$.}
\label{fig:3LoopVacuumDifferent}
\end{figure}
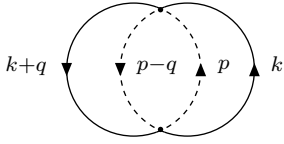

As explained in \cite{HamFur00}, the only nonzero contribution to the
second order correction $E^{(2)}_\Omega$ comes from the Feynman diagram
shown in Figure \ref{fig:3LoopVacuumDifferent}. Performing the same
steps as in the analogous computation \cite{PJ1} of the second order
correction to the ground state energy of a polarized system of spin $1/2$
fermions (to which the present computation reduces in the limit of
$m_a=m_b$) one arrives at the expression
\begin{eqnarray}
  {E^{(2)}_\Omega\over V}={C_0^2\over\hbar}\!\!
  \int\!{d^3\mathbf{q}\over(2\pi)^3}\!\!\int\!{d^3\mathbf{p}\over(2\pi)^3}\!\!
  \int\!{d^3\mathbf{k}\over(2\pi)^3}
{\theta(p_{{\rm F}a}-|\mathbf{k}|)~\!
    \theta(p_{{\rm F}b}-|\mathbf{p}|)~\!
    \theta(|\mathbf{k}+\mathbf{q}|-p_{{\rm F}a})~\!
    \theta(|\mathbf{p}-\mathbf{q}|-p_{{\rm F}b})\over
    \omega^a_{\mathbf{k}}+\omega^b_{\mathbf{p}}-\omega^a_{\mathbf{k}+\mathbf{q}}
    -\omega^b_{\mathbf{p}-\mathbf{q}}+i0}~\!.
  \nonumber
\end{eqnarray}
The next step is passing
to the integrations over the variables $\mathbf{s}$, $\mathbf{t}$ and
$\mathbf{u}$ defined by the relations (the Jacobian equals 8):
\begin{eqnarray}
  \mathbf{k}=\tilde m_a\mathbf{s}-\mathbf{t}~\!,\phantom{aaa}
  \mathbf{p}=\tilde m_b\mathbf{s}+\mathbf{t}~\!,\phantom{aaa}
  \mathbf{q}=\mathbf{t}-\mathbf{u}~\!,\nonumber
\end{eqnarray}
which are the appropriate
modification of those used in \cite{HamFur00,PJ1}, where
\begin{eqnarray}
  \tilde m_{a/b}={2m_{a/b}\over m_a+m_b}~\!,\phantom{aaaa}
  \tilde m_a+\tilde m_b=2~\!. \nonumber
\end{eqnarray}
The denominator of the integrand then becomes equal
$\hbar(\mathbf{t}^2-\mathbf{u}^2+i0)/2m_{\rm red}$ 
and the first and second order corrections to the ground-state energy
can be, after using (\ref{eqn:C0determined}), 
written together in the form
\begin{eqnarray}
  {E^{(1)}_\Omega+E^{(2)}_\Omega\over V}={p^3_{{\rm F}b}p^3_{{\rm F}a}\over9\pi^3}~\!
  {\hbar^2\over2m_{\rm red}}~\!a_0+{2~\!p^3_{{\rm F}b}p^3_{{\rm F}a}\over9\pi^4}~\!
  {\hbar^2\over2m_{\rm red}}~\!a^2_0\Lambda+
  {\hbar^2\over2m_{\rm red}}~\!256a_0^2~\!{\tilde J\over(2\pi)^4}~\!,
  \label{eqn:SumE1andE2}
\end{eqnarray}
where ($\tilde m_b=2-\tilde m_a$)
\begin{eqnarray}
  \tilde J(p_{{\rm F}a},p_{{\rm F}b},\tilde m_a)=\int_0^{s_{\rm max}}\!ds~\!
  s^2{1\over4\pi}\!
  \int\!d^3\mathbf{t}~\!\theta(p_{{\rm F}b}-|\mathbf{t}+\tilde m_b\mathbf{s}|)~\!
  \theta(p_{{\rm F}a}-|\mathbf{t}-\tilde m_a\mathbf{s}|)~\!\tilde g(t,s)~\!,
  \label{eqn:Jpp}\\
  \tilde g(t,s)\equiv\tilde g(|\mathbf{t}|,s)
  ={1\over4\pi}\!\int\!d^3\mathbf{u}~\!
  {\theta(|\mathbf{u}+\tilde m_b\mathbf{s}|-p_{{\rm F}b})~\!
    \theta(|\mathbf{u}-\tilde m_a\mathbf{s}|-p_{{\rm F}a})\over
    \mathbf{t}^2-\mathbf{u}^2+i0}~\!.\phantom{aaaaa}\!\nonumber
\end{eqnarray}
The regions of the integrations over $d^3\mathbf{u}$ and over $d^3\mathbf{t}$
are determined by the intersections of two Fermi spheres of unequal radii,
$p_{{\rm F}b}$ and $p_{{\rm F}a}$, the centers of which are displaced from the
origin of the $\mathbf{u}$ (of the $\mathbf{t}$) space by the vectors
$-\tilde m_b\mathbf{s}$
($\mathbf{s}$ will be taken to determine the $z$-axes of the $\mathbf{u}$ and
$\mathbf{t}$ spaces in the integrals over $d^3\mathbf{u}$ and $d^3\mathbf{t}$)
and $\tilde m_a\mathbf{s}$, respectively (the distance between the centers
is $2|\mathbf{s}|$). This is the only modification compared to the
computation done in \cite{PJ1}. The integral over $\mathbf{u}$ runs over the
infinite exterior of these spheres and is, therefore, divergent; the
integration over $\mathbf{t}$ covers the interior of their intersection.
For this reason the outermost integration over
$s\equiv|\mathbf{s}|$ is restricted
to $s\leq s_{\rm max}={1\over2}(p_{{\rm F}a}+p_{{\rm F}b})$ because if
$s>s_{\rm max}$, the two spheres which determine
the region of the integration over $\mathbf{t}$ become disjoint.

As far as the integral giving $\tilde g(t,s)$ is concerned, the
range of the variable $s$ splits into two domains:
$0\leq s\leq s_0={1\over2}(p_{{\rm F}a}-p_{{\rm F}b})$ and
$s_0\leq s\leq s_{\rm max}$. Correspondingly, the integral $\tilde J$
splits into $\tilde J_1+\tilde J_2$.

If $0\leq s\leq s_0$,
the smaller sphere of radius $p_{{\rm F}b}$ is entirely contained inside the
one of radius $p_{{\rm F}a}$ and plays no role in determining the domain of
integration over $\mathbf{u}$: this domain is then just the (infinite)
exterior of the sphere of radius $p_{{\rm F}a}$ the center of which is at
$u_z=0$, when $s=0$ and moves to the right as $s$ increases.
The computation of $\tilde g(t,s)$ for $0\leq s\leq s_0$ 
and of $\tilde J_1$ proceeds therefore exactly as in the case of equal
masses discussed in \cite{PJ1} and one readily finds that 
\begin{eqnarray}
  \tilde g(t,s)=g(t,~\!\tilde m_as)~\!,\phantom{aaaa}0\leq s\leq s_0~\!,
  \label{eqn:tildetos0}
\end{eqnarray}
where 
\begin{eqnarray}
g(t,s)=-\Lambda+{1\over2}~\!p_{{\rm F}a}
+{t\over4}\ln{(p_{{\rm F}a}-t)^2-s^2\over(p_{{\rm F}a}+t)^2-s^2}
+{p^2_{{\rm F}a}-s^2-t^2\over8s}\ln{(p_{{\rm F}a}+s)^2-t^2\over
  (p_{{\rm F}a}-s)^2-t^2}~\!.\phantom{a}\label{eqn:g2t1}
\end{eqnarray}
is the function obtained in \cite{PJ1} in the case of $m_a=m_b$.

Since $\tilde J_1$ is obtained by integrating the function $\tilde g(t,s)$
first over the interior of the sphere of radius $p_{{\rm F}b}$, the
center of which is shifted by $-\tilde m_b\mathbf{s}$ from the origin
of the $\mathbf{t}$ space, and then, with the weight $s^2$, over $s$
from 0 to $s_0$, it is straightforward to obtain the divergent
part $\tilde J_1^{\rm div}$ of $\tilde J_1$:
\begin{eqnarray}
  \tilde J^{\rm div}_1=-{1\over9}~\!\Lambda~\! s_0^3~\!p^3_{{\rm F}b}=
  -{1\over72}~\!\Lambda~\!(p_{{\rm F}a}-p_{{\rm F}b})^3~\!p^3_{{\rm F}b}~\!.
  \label{eqn:J1div}
\end{eqnarray}
The finite part of $\tilde J_1$ can be easily obtained by numerical
integration. This can be done either by writing
$\mathbf{t}=\mathbf{t}^\prime-\tilde m_b\mathbf{s}$ and introducing the
spherical coordinate system in the $\mathbf{t}^\prime$ space with the
$t^\prime_z$ axis taken in the direction of the vector $\mathbf{s}$:\\
\begin{eqnarray}
\tilde J_1={1\over2}\int_0^{s_0}\!ds~\!s^2\!\int_{-1}^1\!d\eta\int_0^{p_{{\rm F}b}}\!
dt^\prime~\!t^{\prime2}~\!g\!\left(\sqrt{t^{\prime2}-2t^\prime
\tilde m_bs\eta+\tilde m_b^2s^2},~\!\tilde m_as\right),\nonumber
\end{eqnarray}
or just by using the Mathematica instruction
0.5NIntegrate$[s^2t^2g[t,\tilde m_as]~\!{\rm Boole}[t^2+2t\tilde m_bsx
+\tilde m_b^2s^2<p^2_{{\rm F}b}],~\!\{s,0,s_0\},~\!\{x,-1,1\},~\!\{t,0,\infty\}]$.
\vskip0.2cm

\begin{figure}[]
\begin{center}
\begin{picture}(500,120)(5,0)
\CArc(140,50)(50,0,360)
\Vertex(140,50){2}
\CArc(100,50)(30,0,360)
\Vertex(100,50){2}
\Line(65,50)(205,50)
\ArrowLine(205,50)(206,50)
\Line(110,-5)(110,105)
\ArrowLine(110,105)(110,106)
\Text(65,100)[]{$a)$}
\DashLine(110,50)(92,105){3}
\DashCArc(110,50)(15,0,100){1}
\DashArrowArc(110,50)(15,100,101){1}
\Text(117,56)[]{${\vartheta_0}$}
\CArc(340,50)(50,0,360)
\Vertex(340,50){2}
\CArc(300,50)(30,0,360)
\Vertex(300,50){2}
\Line(265,50)(405,50)
\ArrowLine(405,50)(406,50)
\Line(330,-5)(330,105)
\ArrowLine(330,105)(330,106)
\Text(265,100)[]{$b)$}
\DashLine(330,50)(270,110){3}
\DashCArc(330,50)(15,0,125){1}
\DashArrowArc(330,50)(15,125,126){1}
\Text(338,58)[]{${\vartheta_0}$}
\end{picture}
\end{center}
\caption{Intersecting Fermi spheres for
  $(p_{{\rm F}a}-p_{{\rm F}b})/2<s<(p_{{\rm F}a}+p_{{\rm F}b})/2$.
  Dots mark their centers shifted
  by $-\tilde m_bs$ and $\tilde m_as$ from the origin of the space.
  $a)$ $m_a>m_b$, $b)$ $m_a<m_b$. Marked are the
``critical'' polar angles $\vartheta_0$.}
\label{fig:FermiSpheres}
\end{figure}
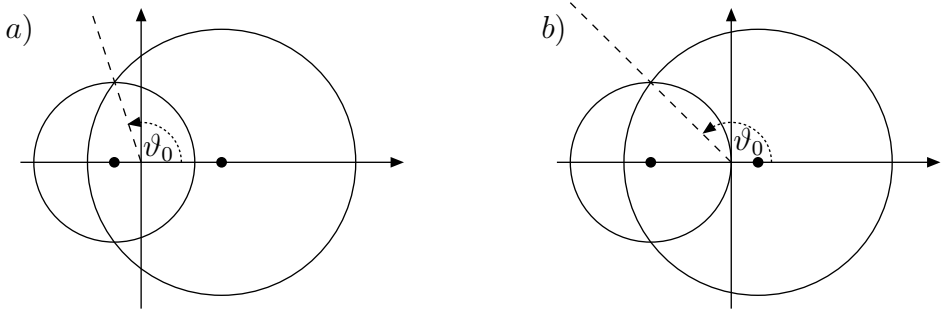

We now compute the function $\tilde g(t,s)$ for
$s_0\leq s\leq s_{\rm max}$ and the corresponding contribution $\tilde J_2$ to
the integral (\ref{eqn:Jpp}).
In this regime the two Fermi spheres which determine the ranges of
integrations over $\mathbf{u}$ and over $\mathbf{t}$ intersect one another.
In the $\mathbf{u}$ space the $z$ coordinate $u_z^0$ of the intersection and
its distance $u_0$ from the origin are determined by  solving the equations
\begin{eqnarray}
  &&u^2_\perp+(u_z-\tilde m_as)^2=p^2_{{\rm F}a}~\!,\nonumber\\
  &&u^2_\perp+(u_z+\tilde m_bs)^2=p^2_{{\rm F}b}~\!,\phantom{aa}\nonumber
\end{eqnarray}
which give (recall that $\tilde m_a+\tilde m_b=2$)
\begin{eqnarray}
  u^0_z=-{1\over4s}\left(p^2_{{\rm F}a}-p^2_{{\rm F}b}\right)
  +{1\over2}(\tilde m_a-\tilde m_b)~\!s~\!,
  \phantom{aaa}u^2_0={1\over2}\left(\tilde m_b
  p^2_{{\rm F}a}+\tilde m_ap^2_{{\rm F}b}\right)-\tilde m_a\tilde m_b~\!s^2~\!.
  \label{eqn:uz0andu0}
\end{eqnarray}
In the spherical system the ``critical'' angles $\vartheta_0$
corresponding to the intersection of the spheres (marked in
Figs. \ref{fig:FermiSpheres}) are given by
\begin{eqnarray}
  \cos\vartheta_0=\xi_0={u_z^0\over u_0}~\!.
\end{eqnarray}
Therefore, if $s_0\leq s\leq s_{\rm max}$ (i.e. when the two Fermi
spheres intersect), the function $\tilde g(t,s)$ is given
by\footnote{Actually this way of computing $\tilde g(t,s)$ in this
  regime ($s_0<s<s_{\rm max}$) is justified geometrically only for $s$
  not greater than some critical value (depending on the ratios
  $p_{{\rm F}b}/p_{{\rm F}a}$ and $m_b/m_a$) which is smaller
  than $s_{\rm max}$. For $s$ greater than critical, the dashed lines
  in Fig. \ref{fig:FermiSpheres} pass through the interiors of the
  smaller spheres and the formula (\ref{eqn:gForIntersectingSpheres})
  may seem to be unjustified. We have checked, however, by integrating
  numerically functions over the domain formed by the exterior of the
  intersecting spheres lying inside a large sphere of radius
  $R>2p_{{\rm F}a}$ that the formula (\ref{eqn:gForIntersectingSpheres})
  always gives the correct answer.}
\begin{eqnarray}
  \tilde g(t,s)=
  {1\over2}\int_{-1}^{\xi_0}\!d\xi\int_{u_b(\xi,s)}^\Lambda\!du~\!{u^2\over t^2-u^2+i0}
  +{1\over2}\int_{\xi_0}^1\!d\xi\int_{u_a(\xi,s)}^\Lambda\!du~\!{u^2\over t^2-u^2+i0}
  ~\!,\label{eqn:gForIntersectingSpheres}
\end{eqnarray}
where 
$u_b(\xi,s)=-\tilde m_bs~\!\xi+\sqrt{p^2_{{\rm F}b}-\tilde m_b^2s^2(1-\xi^2)}$,
$u_a(\xi,s)=\tilde m_as~\!\xi+\sqrt{p^2_{{\rm F}a}-\tilde m_a^2s^2(1-\xi^2)}$;
of course $u_b(\xi_0,s)=u_a(\xi_0,s)\equiv u_0$. After extracting the terms
diverging with $\Lambda\rightarrow\infty$ as in \cite{HamFur00,PJ1} (they
combine to $-2\pi^2I_0$ where $I_0$ is given in (\ref{eqn:I0integralResult}))
one gets
\begin{eqnarray}
  \tilde g(t,s)=-\Lambda-i{\pi\over2}~\!t
  +{1\over2}\int_{-1}^{\xi_0}\!d\xi\int_0^{u_b(\xi,s)}\!du~\!{u^2\over u^2-t^2-i0}
+{1\over2}\int_{\xi_0}^1\!d\xi\int_0^{u_a(\xi,s)}\!du~\!{u^2\over u^2-t^2-i0}
~\!.\nonumber
\end{eqnarray}
It is now straightforward to compute $\tilde J_2^{\rm div}$ and to check
the cancellation of $\Lambda$. Indeed, the integral
\begin{eqnarray}
  {1\over4\pi}\int\!d^3\mathbf{t}~\!
  \theta(p_{{\rm F}b}-|\mathbf{t}+\tilde m_b\mathbf{s}|)~\!
  \theta(p_{{\rm F}a}-|\mathbf{t}-\tilde m_a\mathbf{s}|)
  (-\Lambda)~\!,\nonumber
\end{eqnarray}
can be done by shifting the origin of the
$\mathbf{t}$-space so that the intersection of the two Fermi spheres
occurs at $t_z^\prime=0$. The integration over $d^3\mathbf{t}$
is then easy and its result is
\begin{eqnarray}
  -{\Lambda\over2}\left\{\left[{1\over3}p^3_{{\rm F}b}
    -{1\over2}p^2_{{\rm F}b}(s+u_z^0)
    +{1\over6}(s+u_z^0)^3\right]+
  \left[{1\over3}p^3_{{\rm F}a}-{1\over2}p^2_{{\rm F}a}(s-u_z^0)
    +{1\over6}(s-u_z^0)^3\right]\right\},\nonumber
\end{eqnarray}
where $u_z^0$ is given by (\ref{eqn:uz0andu0}).
This should be integrated from $s_0={1\over2}(p_{{\rm F}a}-p_{{\rm F}b})$
to $s_{\rm max}={1\over2}(p_{{\rm F}a}+p_{{\rm F}b})$ with the weight $s^2$.
Mathematica does the integration readily with the expected result:
\begin{eqnarray}
  \tilde J_2^{\rm div}=-\Lambda\left({p_{{\rm F}a}^2p_{{\rm F}b}^4\over24}
  -{p_{{\rm F}a}p_{{\rm F}b}^5\over24}
  +{p_{{\rm F}b}^6\over72}\right).\nonumber
\end{eqnarray}
Combining this with the divergent part (\ref{eqn:J1div}) of $\tilde J_1$
one gets
\begin{eqnarray}
  \tilde J_1^{\rm div}+\tilde J_2^{\rm div}
  =-\Lambda~\!{p_{{\rm F}b}^3p_{{\rm F}a}^3\over72}~\!,\nonumber
\end{eqnarray}
which is precisely what is needed to cancel in (\ref{eqn:SumE1andE2}) the
term explicitly proportional to $\Lambda$ which comes from expressing
$C_0$ in terms of the scattering length in the first order result. 
\vskip0.2cm

\begin{figure}
\psfig{figure=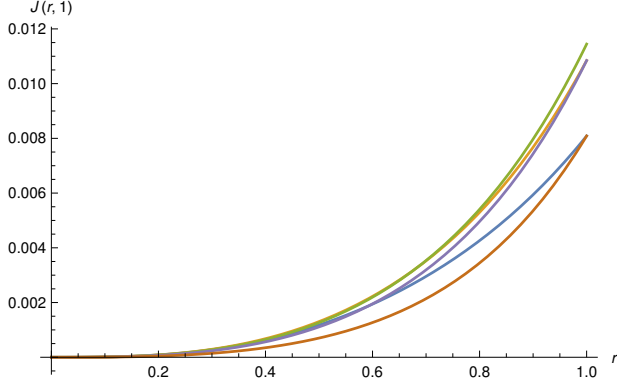,width=8.cm,height=5.0cm} 
\caption{Plot of the function $\tilde J(r,1,\tilde m_a)$. The consecutive
  lines (counting from below at $r\sim0.9$) correspond to the ratio
  $m_b/m_a$ equal to $40/6$ (red), $6/40$ (blue), $2$ (violet), $1/2$
  (yellow) and $1$ (green). A  $r=1$ (zero polarization) the results for
  $m_b/m_a=x$ and $m_b/m_a=1/x$ coincide as they should. The value
  $\tilde J(1,1,1)=0.0114449=(11-2\ln2)/840$ (the endpoint pof the
  green curve for $m_b/m_a=1$) is the the result of \cite{HamFur00}.}
\label{fig:JJplot}
\end{figure}

The remaining integrals in $\tilde g(t,s)$ can be worked out exactly
as in \cite{PJ1} using the trick given in Appendix C of \cite{Mloty2},
that is by taking the integrals over $\xi$ by parts after inserting into
them $1=d\xi/d\xi$. The integrals have imaginary parts which together
precisely cancel the imaginary part which arose from the divergent
integral $I_0$ and the final result for $s_0\leq s\leq s_{\rm max}$ is
\begin{eqnarray}
  \tilde g(t,s)=-\Lambda+{1\over4}(p_{{\rm F}b}+p_{{\rm F}a}+2s)
  +{t\over4}\ln{p_{{\rm F}b}+\tilde m_bs-t\over p_{{\rm F}b}+\tilde m_bs+t}
  +{t\over4}\ln{p_{{\rm F}a}+\tilde m_as-t\over p_{{\rm F}a}+\tilde m_as+t}
  \phantom{aaaaaaaa}\nonumber\\
  +~\!{p^2_{{\rm F}b}-t^2-\tilde m_b^2s^2\over8\tilde m_bs}
  \ln{(p_{{\rm F}b}+\tilde m_bs)^2-t^2\over u_0^2-t^2}
  +{p^2_{{\rm F}a}-t^2-\tilde m_a^2s^2\over8\tilde m_as}
  \ln{(p_{{\rm F}a}+\tilde m_as)^2-t^2\over u_0^2-t^2}~\!,\phantom{a}
 \label{eqn:tildegtosmax}
\end{eqnarray}
where $u_0^2$ is given in (\ref{eqn:uz0andu0}).
In the limit  $m_a=m_b$ (i.e. $\tilde m_a=\tilde m_b=1$) the results
(\ref{eqn:tildetos0}) and (\ref{eqn:tildegtosmax}) go over into
the ones obtained in \cite{PJ1} which agree with the result obtained
in \cite{KANNO}. The finite parts of the functions $\tilde J_1$ and
$\tilde J_2$, i.e. the integrals over $t$, $\eta$ and $s$, can be easily
evaluated using for instance the standard Mathematica function allowing
to numerically perform integrations over (multidimensional) domains.
Since the finite part of $\tilde J=\tilde J_1+\tilde J_2$ scales
as the seventh power of $p_{{\rm F}a}$, in Fig. \ref{fig:JJplot} we show
$\tilde J(r,1,\tilde m_a)$ as a function of $r=p_{{\rm F}b}/p_{{\rm F}a}$
for several values of the mass ratio $m_b/m_a$. It is clear that
the curves corresponding to $m_b/m_a=x$ and $1/x$ merge for $r=1$
(vanishing polarization) as they should; also, independently of
the mass ratio, the function $\tilde J$ vanishes for the
maximal polarization (at $P=1$, i.e. for  $r=0$) when all fermions are
of the same type, as required by the Pauli exclusion principle.

\section{Results}
\label{sec:res}

The energy density with the order $(k_{\rm F}a_0)^2$ term included
can be expressed in several equivalent ways. Either in terms of the
ratio $r\equiv p_{{\rm F}b}/p_{{\rm F}a}$ and
$k_{\rm F}\equiv3\pi^2(N_a+N_b)/V$, so that $p_{{\rm F}a}=k_{\rm F}(2/(1+r^3))^{1/3}$,
\begin{eqnarray}
  {E_\Omega\over V}={k_{\rm F}^3\over3\pi^2}~\!{\hbar^2k_{\rm F}^2\over4m_{\rm red}}
  \left({2\over1+r^3}\right)^{5/3}
  \left\{{3\over10}\left(\tilde m_b+\tilde m_ar^5\right)+{2\over3\pi}~\!
  r^3\left({2\over1+r^3}\right)^{1/3}(k_{\rm F}a_0)\right.\phantom{aa}\nonumber\\
  \left.  +{96\over\pi^2}\left({2\over1+r^3}\right)^{2/3}(k_{\rm F}a_0)^2
  \tilde J(1,r,\tilde m_a)+\dots\right\},~
\end{eqnarray}
or in terms of the polarization $P=(N_a-N_b)/(N_a+N_b)=(1-r^3)/(1+r^3)$
\begin{eqnarray}
  {E_\Omega\over V}={k_{\rm F}^3\over3\pi^2}~\!{\hbar^2k_{\rm F}^2\over4m_{\rm red}}
  \left\{{3\over10}\left(\tilde m_b(1+P)^{5/3}+\tilde m_a(1-P)^{5/3}\right)
  +{2\over3\pi}~\!(1-P^2)~\!(k_{\rm F}a_0)\right.\nonumber\\
  \left.+{96\over\pi^2}\left(1+P\right)^{7/3}(k_{\rm F}a_0)^2
  \tilde J(1,r(P),\tilde m_a)+\dots\right\},\phantom{aaaaaaaaa}
  \label{eqn:Final}
\end{eqnarray}
where $r(P)=((1-P)/(1+P))^{1/3}$. Note also that the prefactor
$\hbar^2k^5_{\rm F}/12\pi^2m_{\rm red}$ in this formula can  be written in the
form $(N/V)(\hbar^2k^2_{\rm F}/4m_{\rm red})$. At zero
polarization ($r=1$, $P=0$) the formula (\ref{eqn:Final}) simplifies to
\begin{eqnarray}
{E_\Omega\over V}={k_{\rm F}^3\over3\pi^2}~\!
{\hbar^2k_{\rm F}^2\over4m_{\rm red}}
{3\over5}\left\{1+{10\over9\pi}~\!(k_{\rm F}a_0)
+{160\over\pi^2}~\!(k_{\rm F}a_0)^2\tilde J(1,1,\tilde m_a)+\dots\right\}.
\end{eqnarray}
Setting here $\tilde m_a=40/23$, i.e. $m_b/m_a=6/40$, one finds that
with $\tilde J=0.00808856$ this
agrees with the second order result shown for this mass ration in Fig.~1
of \cite{FraPil}. No plots of energy density for nonzero
polarization are shown in \cite{FraPil} but the authors
give an interpolation formula for the evaluated numerically
function $I(P,~\!m_b/m_a)$ in terms of which their second order
correction to the system's energy is expressed. The precise relation
of this function $I$ to our function $\tilde J$ should be
\begin{eqnarray}
  {m_a+m_b\over m_b}~\!I(P,~\!m_b/m_a)=320~\!(1+P)^{7/3}~\!
  \tilde J(1,~\!((1-P)/(1+P))^{1/3},~\!2m_a/(m_a+m_b))~\!.\nonumber
\end{eqnarray}
We have checked that although the interpolation formula of
\cite{FraPil} yields for $P=1$ a small but nonzero value of the function
$I$ (slightly at variance with the Pauli exclusion principle), it
nevertheless agrees excellently with the results of our calculation.

\begin{figure}
\psfig{figure=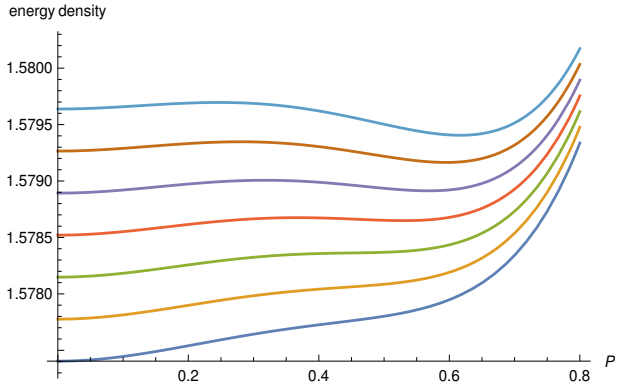,width=8.cm,height=5.0cm} 
\caption{Energy density $E_\Omega/V$ in units
  $(3/5)(\hbar^2k^2_{\rm F}/2m_f)(k^3_{\rm F}/3\pi^2)$ of the gas of
  same mass ($m_a=m_b\equiv m_f$) spin $1/2$ fermions as a function of
  the polarization $P$
  for different values (from below): 1.0520 (blue), 1.0525 (yellow),
  1.0530 (green), 1.0535 (red), 1.0540 (violet), 1.0545 (dark red)
  and 1.0550 (light blue) of $k_{\rm F}a_0$. The emergence of
  the ferromagnetic behaviour as well as the first order character
  of the transition to the ferromagnetic state are clearly seen.}
\label{fig:PhTr}
\end{figure}

The energy density given by the formula (\ref{eqn:Final}) in the case
of equal masses ($m_a=m_b\equiv m_f$) of the oppositely polarized fermions
is shown in Figure \ref{fig:PhTr}. It illustrates the well-known fact
\cite{DuiMac}
that the emergence of a nonzero polarization (of the global minimum
of the energy density as a function of $P$) which in the mean-field
approximation (i.e. with only the order $k_{\rm F}a_0$ term included)
is a second order transition \cite{Sto}, after the inclusion of the second
order term becomes the first order one. This, however, occurs probably
beyond the limits of the reliability of the approximation used:
for vanishing polarization the comparison of the second order formula 
with the results of the quantum Monte Carlo simulations of \cite{QMC10}
shows that it is numerically reliable only up to $k_{\rm F}a_0\simlt0.5$.

\begin{figure}
\psfig{figure=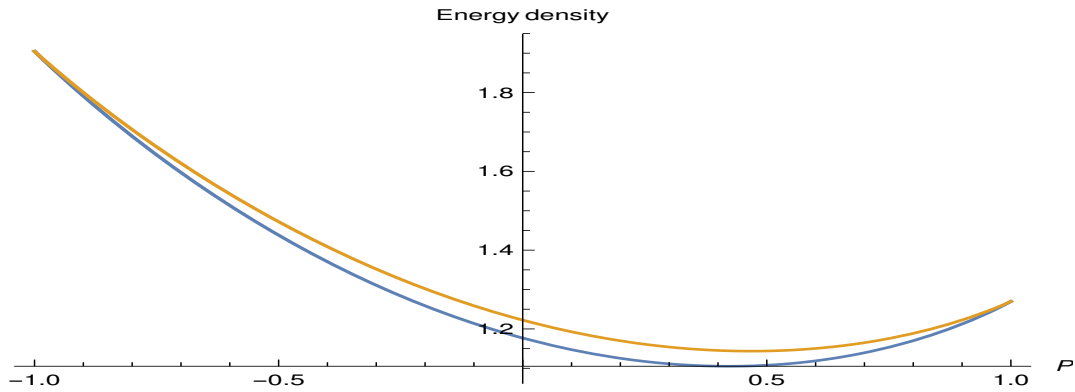,width=14.cm,height=5.0cm} 
\caption{Energy density $E_\Omega/V$ in units 
  $(3/5)(\hbar^2k^2_{\rm F}/4m_{\rm red})(k^3_{\rm F}/3\pi^2)$
  of the mixture of two species of oppositely polarized 
  spin $1/2$ fermions with the mass ratio $m_b/m_a=2/3$ and
  $k_{\rm F}a_0=0.5$,
  as a function of the polarization $P$. The lower curve is the
  mean-field result, while the upper one shows the effects of
inclusion the term of order $(k_{\rm F}a_0)^2$.}
\label{fig:MassImb}
\end{figure}

If the masses of oppositely polarized fermions differ, the minimum of 
the energy density is at $P\neq0$ already in the case of vanishing
interactions (the system is polarized in the direction of the
polarization of the heavier species). If the mass ratio corresponds
to different atoms, this effects completely dominates the dependence
of the energy density on the polarization. Only if the mass ratio
is very close to unity (as would be if different isotopes of the
same element could play the roles two different fermion species),
can the mean field correction generate a higher (i.e. unstable),
second minimum at an opposite polarization and this only when
$k_{\rm F}a_0\approx\pi/2$; for such strengths of the interaction,
however, the second order correction computed in this paper and in
\cite{FraPil} is
so large, that the two minima (the deeper one corresponding to
the direction of the polarization of heavier isotopes) occur already
at $P=\pm1$ and the discussion of the change of the order of the
transition is meaningless in view of the clear unreliability of
the expansion. In Fig. \ref{fig:MassImb}
the effects of inclusion of the second order term in the
case of the mass ratio $m_b/m_a=2/3$ are shown for $k_{\rm F}a_0=0.5$
to show that when the expansion is reliable, the minimum of the
energy density at a nonzero polarization, existing already without
any interaction, can only be slightly displaced.

\section{Final remarks}
\label{sec:summary}

We have shown that the order $(k_{\rm F}a_0)^2$, where $a_0$ is the $s$-wave
scattering length and $k_{\rm F}=(3\pi^2N/V)^{1/3}$, correction to the
ground-state energy of a mixture of oppositely polarized fermions of
different masses, computed for the first time in \cite{FraPil}
by using a complicated numerical evaluation of the formulae derived
in \cite{AbGoDz}, can be easily and semianalytically reproduced
in the effective theory approach proposed first in \cite{HamFur00}
which is simple and leads to integrals which can be numerically evaluated
using the standard built-in Mathematica routines.
Thus this method can allow for extending the computation to yet higher
orders.
\vskip0.3cm

\noindent{\bf Acknowledgments.} We would like to thank Sebastiano Pilati
and the anonymous referee of the paper \cite{PJ1}
for suggesting us this extension of the method. 


\vskip0.5cm

\begin{thebibliography}{99}
\bibitem{FraPil} E. Fratini, S. Pilati, {\em Phys. Rev.}
  {\bf A90}, 023605 (2014).
\bibitem{Jo} G.-B. Jo, Y.-R. Lee, J.-H. Choi, C.A. Christensen, T.H. Kim,
  J.H.Thywissen, D.E. Pritchard and W. Ketterle, {\em Science} {\bf 325} (2009)
  1521.
\bibitem{Lee} Y.-R. Lee, M.-S. Heo, J.-H. Choi, T.T. Wang, C.A. Christensen,
  T.M. Rvachov and W. Ketterle, {\em Phys. Rev.} {\bf A85} (2012), 063615;
  C. Sanner, E.J. Su, W. Huang, A. Keshet, J. Gillen and W. Ketterle,
  {\em Phys. Rev. Lett.} {\bf108} (2012), 240404.
\bibitem{Sto} E. Stoner, {\em Philos. Mag.} {\bf 15}, 1018 (1933); see also
  Section 13.4 in K. Huang, {\it Statistical Mechanics}, John Willey and
  Sons, Inc., New York 1963.
\bibitem{Cond} G.J. Conduit, A.G. Green and B.D. Simons,
  {\em Phys. Rev. Lett.} {\bf 103}, 207201 (2009); S.-Y. Chang, M. Randeria
  and N. Trivedi, {\em Proc. Natl. Acad. Sci. USA} {\bf 108}, 51 (2010)
\bibitem{QMC10} S. Pilati, G. Bertaina, S. Giorgini and M. Troyer,
  {\em Phys. Rev. Lett.} {\bf 105}, 030405 (2010),
  {\sf arXiv:1004/1169 [cond-mat.quant-gas]}.
\bibitem{Pek} D. Pekker, M. Babadi, R. Sensarma, N.  Zinner, L. Polleti,
  W.M. Zwierlein and E. Demler, {\em Phys. Rev. Lett.} {\bf 106},
  050402 (2011).
\bibitem{AbGoDz} A.A. Abrikosov, L.P. Gorkov and I.E. Dzialoshinski,
  {\it Methods of Quantum Field Theory in Statistical Physics}, Dover
  Publications, Inc. New York, 1975.
\bibitem{HamFur00} H.-W. Hammer, R. J. Furnstahl, {\em Nucl. Phys.} A {\bf 678},
  277 (2000); {\sf arXiv:nucl-th/0004043}.
\bibitem{HAKOL} H.-W. Hammer, S. K\"onig and U. van Kolck,
  {\em Rev. of Mod. Phys.} {\bf 92} (2020), 025004.
\bibitem{KolckiSka} see e.g. Proceedings of the Joint Caltech/INT Workshop
  {\it Nuclear Physics with Effective Field Theory}, ed. R. Seki, U. van Kolck
  and M. Savage (World Scientific, 1998); Proceedings of the INT Workshop
  {\it Nuclear Physics with Effective Field Theory II}, ed. P.F. Bedaque,
   M. Savage, R. Seki and U. van Kolck (World Scientific, 2000).
\bibitem{WeDrSch} C. Wellenhofer, C. Drischler and A. Schwenk,
  {\em Phys. Lett. B} {\bf 802} (2020) 135247.
\bibitem{PJ1} P. Chankowski and J. Wojtkiewicz, {\em Phys. Rev.}
  {\bf B104}, 144425 (2021).
\bibitem{KANNO} S. Kanno, {\em Prog. Theor. Phys.} {\bf 44}, 813 (1970).
\bibitem{Mloty2} R. J. Furnstahl, H.-W. Hammer and N. Tirfessa,
  {\em Nucl. Phys. A} {\bf 689},  846 (2001).
\bibitem{FetWal} A. L. Fetter and J. D. Walecka, Quantum Theory of Many Particle
  Systems. McGraw Hill, 1971.
\bibitem{Feynman} R.P. Feynman, {\it Statistical Mechanics. A Set of Lectures},
  W.A. Benjamin, Inc. 1972.
\bibitem{Weinb} S. Weinberg, {\em The Quantum Theory of Fields}, Vol. I,
  The Press Syndicate of the University of Cambridge, 1995.
\bibitem{DuiMac} R.A. Duine, A.H. MacDonald, {\em Phys. Rev. Lett.}
  {\bf 95}, 230403 (2005).
\end{thebibliography}
\end{document}